 \documentclass[final,3p,times,twocolumn]{elsarticle}

\usepackage{amssymb}
\usepackage{amsmath}

\usepackage{hyperref}
\usepackage{graphicx,tikz,pgf}
\usepackage[dvipsnames]{xcolor}
\usetikzlibrary{cd,shapes,decorations.pathmorphing,decorations.markings,decorations.pathreplacing}
\usepackage{ifthen,xparse}
\biboptions{sort&compress}

\definecolor{BlobLineColour}{RGB}{96, 96, 96} 
\definecolor{BlobFillColour}{RGB}{96, 96, 96}  
\definecolor{GluonColour}{RGB}{64, 64, 64} 
\definecolor{DoubleGluonColour}{RGB}{64, 64, 64} 
\definecolor{GhostColour}{RGB}{64, 64, 64} 
\definecolor{QuarkColour}{RGB}{64, 64, 64} 
\definecolor{PlainColour}{RGB}{64, 64, 64} 
\definecolor{DoublePlainColour}{RGB}{64, 64, 64} 
\definecolor{ChargedScalarColour}{RGB}{64, 64, 64} 
\definecolor{NeutralScalarColour}{RGB}{64, 64, 64}

\def\picSc{0.55} 
\def\blobSc{0.5} 
\def\bigBlobSc{0.75}

\def\lineW{0.5} 

\ExplSyntaxOn
\NewDocumentCommand{\newwhiledo}{m m}
  {
   \bool_while_do:nn { \int_compare_p:n {#1} } { #2 }
  }
\ExplSyntaxOff

\newcommand\blobNode[4]{\node[circle,line width=\lineW mm,BlobLineColour!80,draw,fill=BlobFillColour!10] at (#1,#2)[scale=#3] {#4};}

\newcommand\leftArc[2]{
	\pgfmathsetmacro{\np}{#1} 
	\ifthenelse{\equal{\np}{1}}	{
		\def\angle{180}
		\node[circle,line width=\lineW mm,BlobLineColour!80,draw,fill=BlobFilleColour!10] (V\np) at (\angle: 1)[scale=\blobSc] {#2};
	}	{
		\foreach \k in {1,...,\np}
			\def\angle{{ 360 * ( \k / (2*\np+2) ) + 90 }}
			\node[circle,line width=\lineW mm,BlobLineColour!80,draw,fill=BlobFillColour!10] (V\k) at (\angle: 1)[scale=\blobSc] {#2};
			\draw [line width=\lineW mm]  \foreach \x [remember=\x as \lastx (initially 1)] in {2,...,\np}{(V\lastx) to [bend right] (V\x)};
	}
	\draw [line width=\lineW mm] (V1) to [bend left] (0,1);
	\draw [line width=\lineW mm] (V\np) to [bend right] (0,-1);
}

\newcommand\rightArc[2]{
	\pgfmathsetmacro{\np}{#1} 
	\ifthenelse{\equal{\np}{1}}	{
		\def\angle{0}
		\node[circle,line width=\lineW mm,BlobLineColour!80,draw,fill=BlobFillColour!10] (V\np) at (\angle: 1)[scale=\blobSc] {#2};
	}	{
		\foreach \k in {1,...,\np}
			\def\angle{{ -1 * 360 * ( \k / (2*\np+2) ) + 90 }}
			\node[circle,line width=\lineW mm,BlobLineColour!80,draw,fill=BlobFillColour!10] (V\k) at (\angle: 1)[scale=\blobSc] {#2};
			\draw [line width=\lineW mm] \foreach \x [remember=\x as \lastx (initially 1)] in {2,...,\np}{(V\lastx) to [bend left] (V\x)};
	}
	\draw [line width=\lineW mm] (V1) to [bend right] (0,1);
	\draw [line width=\lineW mm] (V\np) to [bend left] (0,-1);
}

\newcommand\gluonLine[4]{
	\draw[decorate, decoration={snake,amplitude=.4mm,segment length=2mm,post length=0mm}, line width=\lineW mm, GluonColour!100] (#1,#2) -- (#3,#4);
}

\newcommand\gluonLoop[3]{
	\draw[decorate, decoration={snake,amplitude=.4mm,segment length=2mm,post length=0mm}, line width=\lineW mm, GluonColour!100,fill=White] (#1,#2) circle (#3);
}

\newcommand\ghostLoop[5]{
	\tikzset{deco/.style n args={4}{decoration={markings, mark=at position ##1 with { \draw [<-] (0,0) --  (3pt,0)node [near end,##2=5 pt]{##3};}, mark=at position ##4 with { \draw [<-] (0,0) --  (3pt,0)node [near end,##2=5 pt]{##3};}},  postaction={decorate}}};
	\draw[deco={#4}{left}{}{#5},loosely dotted, line width=\lineW mm, GhostColour!100,->>,fill=White] (#1,#2) circle (#3);
}

\newcommand\quarkLoop[5]{
	\tikzset{deco/.style n args={4}{decoration={markings, mark=at position ##1 with { \draw [<-] (0,0) --  (3pt,0)node [near end,##2=5 pt]{##3};}, mark=at position ##4 with { \draw [<-] (0,0) --  (3pt,0)node [near end,##2=5 pt]{##3};}},  postaction={decorate}}};
	\draw[deco={#4}{left}{}{#5}, line width=\lineW mm, QuarkColour!100,->>,fill=White] (#1,#2) circle (#3);
}

\newcommand\doublePlainLoop[3]{
	\draw[style=double, line width=\lineW mm, DoublePlainColour!100,fill=White] (#1,#2) circle (#3);
}

\newcommand\gluonArc[5]{
	\draw[decorate, decoration={snake,amplitude=.4mm,segment length=2mm,post length=0mm}, line width=\lineW mm, GluonColour!100] ({#1+#3*cos(#4)},{#2+#3*sin(#4)}) arc (#4:#5:#3);
}

\DeclareMathOperator*{\sumint}{                                      \mathchoice
  {\ooalign{$\displaystyle\sum$\cr\hidewidth$\displaystyle\int$\hidewidth\cr}}
  {\ooalign{\raisebox{.14\height}{\scalebox{.7}{$\textstyle\sum$}}\cr\hidewidth$\textstyle\int$\hidewidth\cr}}
  {\ooalign{\raisebox{.2\height}{\scalebox{.6}{$\scriptstyle\sum$}}\cr$\scriptstyle\int$\cr}}
  {\ooalign{\raisebox{.2\height}{\scalebox{.6}{$\scriptstyle\sum$}}\cr$\scriptstyle\int$\cr}}
  }

\begin{document}

\begin{frontmatter}

\title{Perturbative study of NLO chromoelectric correlators in Euclidean space}

\author[a,b]{Panayiotis Panayiotou}
\affiliation[a]{
organization={Technical University of Munich, TUM School of Natural Sciences, Physics Department},   
addressline={James-Franck-Str. 1},city={Garching}, postcode={85748} , country={Germany}}
\affiliation[b]{organization={Excellence Cluster ORIGINS}, addressline={
Boltzmannstrasse 2}, city={Garching}, postcode={85748}, country={Germany}}
\begin{abstract}
We report on the perturbative study, at next-to-leading order (NLO), of  correlation functions at finite temperature of two chromoelectric fields connected by an adjoint Wilson line in Euclidean space. We find a source of asymmetry in two of the correlators studied. Finally, we compare the results with recent Lattice QCD calculations at high temperatures and show a good agreement between the two.
\end{abstract}


\end{frontmatter}


\section{Introduction}
\label{Introduction}
Heavy quarkonium is a central probe of the quark–gluon plasma (QGP), created in relativistic heavy-ion collisions. Because of the short lifetime of the QGP, quarkonium suppression provides an indirect but clean signal of the existence of the medium, as originally proposed by Matsui and Satz \cite{Matsui:1986dk}. Over the past decades, heavy-ion experiments at RHIC and the LHC have provided data on quarkonium suppression \cite{PHENIX:2011img,ALICE:2012jsl,CMS:2016rpc,ATLAS:2018hqe,STAR:2019fge}, suggesting an interplay between different in-medium mechanisms such as colour screening \cite{Karsch:1987pv}, thermal dissociation \cite{Kharzeev:1994pz,Xu:1995eb,Grandchamp:2001pf,Grandchamp:2002wp,Laine:2006ns,Brambilla:2008cx,Brambilla:2011sg,Brambilla:2013dpa}, and regeneration \cite{Braun-Munzinger:2000csl,Thews:2000rj,Andronic:2007bi,Zhao:2007hh,Brambilla:2023hkw}.
Within pNRQCD at finite temperature \cite{Brambilla:2008cx,Escobedo:2008sy,Brambilla:2010vq}, the real-time propagation of quarkonium inside the medium can be formulated as an open quantum system problem, where the QGP plays the role of the environment. In this framework, the evolution equations of the quarkonium density matrix are governed by transport coefficients that can be expressed in terms of gauge-invariant correlators of chromoelectric fields connected by adjoint Wilson lines \cite{Brambilla:2016wgg}.

These chromoelectric correlators are the focus of these proceedings, which are based on \cite{Brambilla:2025xnw}. Previous studies have mostly addressed the real-time correlators associated with the quarkonium evolution or dealt with the extraction of transport coefficients related to the propagation of a heavy quark in the plasma, which can be related to a correlator of two chromoelectric fields connected by fundamental Wilson lines \cite{Binder:2021otw,Biondini:2023zcz,Eller:2019spw,Scheihing-Hitschfeld:2023tuz,Caron-Huot:2009ncn,Burnier:2010rp}. Recently, a first lattice calculation of the adjoint correlators has appeared \cite{Brambilla:2025cqy}, revealing in particular that the correlator associated with singlet–octet transition in pNRQCD exhibits an asymmetry along the thermal circle.

In these conference proceedings, we report on the computation of the three gauge-invariant correlators constructed from two chromoelectric fields connected by adjoint Wilson lines at finite temperature using imaginary-time formalism  to next-to-leading order (NLO) in the weak coupling, making use of integration-by-parts (IBP) techniques. This is supplemented by the leading soft contributions associated with the Debye scale, making the results accurate up to order $g_s^5$. We show that the results are consistent with the lattice calculation in \cite{Brambilla:2025cqy}.
\section{The adjoint chromoelectric correlators}
\label{Correlators}
In Euclidean space, we find that there are only three distinct and gauge-invariant correlators with two chromoelectric fields, $E^a_i(t)$, connected with an adjoint Wilson line. These are,
\begin{align}
    \left\langle EE \right\rangle_U \equiv & -\left\langle g_s E^a_{i}(0) U^{ab}(0,t) g_s E^b_{i} (t) \right\rangle T^{-4},\label{defEEU}\\
    \left\langle EE \right\rangle_L \equiv & -\left\langle g_s E^b_{i}(0) g_s E^a_{i} (t) U^{ab}(t,1/T) \right\rangle T^{-4}, \label{defEEL} \\
    \left\langle EE \right\rangle_S \equiv & -\left\langle g_s E_i^{ab}(0) U^{bc}(0,t)g_s E_i^{cd}(t) U^{da}(t,1/T) \right\rangle T^{-4}, \label{defEES}
\end{align}
where we call $\left\langle EE \right\rangle_U$, $\left\langle EE \right\rangle_L$ and $\left\langle EE \right\rangle_S$, the upper, lower and symmetric correlator respectively. $U^{ab}(t^{\prime},t)$  denotes the Wilson line in an adjoint representation connecting the two points $t^{\prime}$ and $t$ on the thermal circle. Specifically the Wilson line in the adjoint representation reads,
\begin{align}
    U^{ab}(t,t') \equiv \left[\mathrm{P} \exp\left( -i g_s  \int_t^{t'} \mathrm{d}t''
    A_0(t'')  \right)\right]^{ab},
\end{align}
where $(A_\mu)_{ab}=(A_\mu^cT_c)_{ab}$ is the gauge field in the adjoint representation of $\mathrm{SU}(N_c)$ with $\left(T_c\right)_{ab}=-if_{abc}$, and P stands for path ordering.
In the symmetric correlator, chromoelectric fields are constructed as matrices in colour space with the relation $E^{ab}_{i}(t)\equiv x_{abc} E^c_i (t)$ and $x_{abc} \in \{f_{abc},d_{abc}\}$, $(f_{abc})\,d_{abc}$ being the (anti-)symmetric structure constants of $\mathrm{SU}(N_c)$, connecting them with the conventional chromoelectric fields. Finally, we choose to multiply the correlation functions by $-T^{-4}$ to render them dimensionless and positive. 

Each correlator has a different physical interpretation. In \cite{Brambilla:2016wgg,Brambilla:2017zei, Scheihing-Hitschfeld:2023tuz}, the authors have shown that the correlator in eq. \eqref{defEEU} is connected to quarkonium dissociation in pNRQCD inside the medium and the real-time correlator associated with eq. \eqref{defEEL} is responsible for recombination of quarkonium inside the QGP \cite{Brambilla:QuarkoniumTransportCoefficients}. The symmetric correlator has a different interpretation depending on the structure constant we choose to study. For example, for the symmetric structure constant, the symmetric correlator encodes the octet-to-octet transition in pNRQCD, whereas for the antisymmetric structure constant, the correlator, although not clear, can be related to an adjoint open heavy quark diffusing inside QGP \cite{Brambilla:2025xnw}. Related works on the correlator associated with heavy quark diffusion can be found in \cite{Burnier:2010rp}.

\section{Methods}
\label{Methods}
Using standard Euclidean Feynman rules, found for example in \cite{Bellac:2011kqa}, we construct diagrams that contribute at NLO. The nine distinct non vanishing topologies, shared in all three correlators are:
\begin{align}\label{eq: diagrams}
\resizebox{0.2\columnwidth}{!}{
\begin{tikzpicture}[scale=\picSc]
	\doublePlainLoop{2}{0}{2}
	\gluonLine{0}{0}{4}{0}
	\blobNode{0}{0}{\bigBlobSc}{$E$}
	\blobNode{4}{0}{\bigBlobSc}{$E$},
	\gluonLoop{2}{-0.2}{0.70}
\end{tikzpicture}}&  
\resizebox{0.2\columnwidth}{!}{
 \begin{tikzpicture}[scale=\picSc]
 	\doublePlainLoop{2}{0}{2}
 	\gluonLine{0}{0}{4}{0}
 	\blobNode{0}{0}{\bigBlobSc}{$E$}
 	\blobNode{4}{0}{\bigBlobSc}{$E$},
 	\gluonLoop{2}{+0.3}{0.7}
 \end{tikzpicture}} 
 \resizebox{0.2\columnwidth}{!}{
 \begin{tikzpicture}[scale=\picSc]
 	\doublePlainLoop{2}{0}{2}
 	\gluonLine{0}{0}{4}{0}
 	\blobNode{0}{0}{\bigBlobSc}{$E$}
 	\blobNode{4}{0}{\bigBlobSc}{$E$},
 	\ghostLoop{2}{-0.1}{0.7}{0.25}{0.75}
 \end{tikzpicture}}  
 \resizebox{0.2\columnwidth}{!}{
 \begin{tikzpicture}[scale=\picSc]
 	\doublePlainLoop{2}{0}{2}
 	\gluonLine{0}{0}{4}{0}
 	\blobNode{0}{0}{\bigBlobSc}{$E$}
 	\blobNode{4}{0}{\bigBlobSc}{$E$},
 	\quarkLoop{2}{-0.1}{0.8}{0.25}{0.75}
 \end{tikzpicture}}\nonumber\\
     &\resizebox{0.2\columnwidth}{!}{
 \begin{tikzpicture}[scale=\picSc]
 	\doublePlainLoop{2}{0}{2}
 	\gluonArc{2}{-2.2}{3}{40}{140}
 	\gluonArc{2}{+2.2}{3}{-40}{-140}
 	\blobNode{0}{0}{\bigBlobSc}{$E$}
 	\blobNode{4}{0}{\bigBlobSc}{$E$} 
 \end{tikzpicture}}
 \resizebox{0.2\columnwidth}{!}{
 \begin{tikzpicture}[scale=\picSc]
 	\doublePlainLoop{2}{0}{2}
 	\gluonLoop{1.25}{-0.1}{0.8}
 	\gluonLine{2.1}{0}{4}{0}
 	\blobNode{0}{0}{\bigBlobSc}{$E$}
 	\blobNode{4}{0}{\bigBlobSc}{$E$}
 \end{tikzpicture}} 
\resizebox{0.2\columnwidth}{!}{
 \begin{tikzpicture}[scale=\picSc]
 	\doublePlainLoop{2}{0}{2}
 	\gluonLoop{2.8}{-0.1}{0.8}
 	\gluonLine{0}{0}{1.9}{0}
 	\blobNode{0}{0}{\bigBlobSc}{$E$}
 	\blobNode{4}{0}{\bigBlobSc}{$E$}
 \end{tikzpicture}}\nonumber\\
    &\resizebox{0.2\columnwidth}{!}{
 \begin{tikzpicture}[scale=\picSc]
 	\doublePlainLoop{2}{0}{2}
 	\gluonLine{0}{0}{4}{0}
 	\gluonArc{0}{2}{1.5}{-5}{-90}
 	\blobNode{0}{0}{\bigBlobSc}{$E$}
 	\blobNode{4}{0}{\bigBlobSc}{$E$} 
 \end{tikzpicture}} 
 \resizebox{0.2\columnwidth}{!}{
 \begin{tikzpicture}[scale=\picSc]
 	\doublePlainLoop{2}{0}{2}
 	\gluonLine{0}{0}{4}{0}
 	\gluonArc{4}{2}{2}{185}{270}
 	\blobNode{0}{0}{\bigBlobSc}{$E$}
 	\blobNode{4}{0}{\bigBlobSc}{$E$} 
 \end{tikzpicture}} 
\resizebox{0.2\columnwidth}{!}{
 \begin{tikzpicture}[scale=\picSc]
 	\doublePlainLoop{2}{0}{2}
 	\gluonLine{0}{0}{4}{0}
 	\gluonLine{2}{1.9}{2}{0}
 	\blobNode{0}{0}{\bigBlobSc}{$E$}
 	\blobNode{4}{0}{\bigBlobSc}{$E$}
 \end{tikzpicture}} 
\end{align}
There are two further topologies needed for the evaluation of the symmetric correlator. These are:
\begin{equation}\label{eq: symdiagrams}
\resizebox{0.2\columnwidth}{!}{
\begin{tikzpicture}[scale=\picSc]
	\doublePlainLoop{2}{0}{2}
	\gluonArc{0}{2}{2}{-5}{-90}
	\gluonArc{4}{-2}{2}{90}{175}
	\blobNode{0}{0}{\bigBlobSc}{$E$}
	\blobNode{4}{0}{\bigBlobSc}{$E$},
\end{tikzpicture}} \quad
\resizebox{0.2\columnwidth}{!}{
\begin{tikzpicture}[scale=\picSc]
	\doublePlainLoop{2}{0}{2}
	\gluonArc{0}{-2}{2}{0}{90}
	\gluonArc{4}{+2}{2}{180}{270}
	\blobNode{0}{0}{\bigBlobSc}{$E$}
	\blobNode{4}{0}{\bigBlobSc}{$E$},
\end{tikzpicture}} \quad
\resizebox{0.2\columnwidth}{!}{
\begin{tikzpicture}[scale=\picSc]
	\doublePlainLoop{2}{0}{2}
	\gluonLine{0}{0}{4}{0}
	\gluonLine{2}{1.9}{2}{0.3}
	\gluonLine{2}{-0.3}{2}{-1.9}
	\blobNode{0}{0}{\bigBlobSc}{$E$}
	\blobNode{4}{0}{\bigBlobSc}{$E$}
\end{tikzpicture}} \,\,
\end{equation}

The diagrams are understood as follows: The chromoelectric field insertions are the nodes of each diagram, which denote the Euclidean time $0$ and $t$ respectively. The double line wrapping around the chromoelectric field insertions denotes the Wilson lines. Although this graphical notation is only valid for correlators with two Wilson lines, if we impose that an adjoint Wilson line without internal lines connecting to it is diagrammatically equivalent to a colour-space $\delta_{ab}$, then with this observation, we obtain a consistent graphical representation for all three correlators. Note that here we have drawn only the diagrams that contribute to the upper correlator. Contributions to the other correlators are obtained by trivially reflecting the diagrams on a horizontal axis. 

At NLO, we write down the diagrams using conventional Euclidean QCD Feynman rules, perform colour and Lorentz contractions, scalarizing the two-loop integrals we encounter. At this point, we employ an integration-by-parts identity \cite{Davydychev:2022dcw,Davydychev:2023jto} for the spatial integrals (in this case, only one is required),
\begin{align}
\int_{\mathbf{pq}} &\frac{1}{(p_0^2+p^2)(q_0^2+q^2)\left[(p_0+q_0)^2+(\mathbf{p}+\mathbf{q})^2\right]} \nonumber\\
&= -\frac{d-2}{2\left(d-3\right)}
\Bigg[\frac{I_{1}\left(p_{0}\right)I_{1}\left(p_{0}+q_{0}\right)}{p_{0}\left(p_{0}+q_{0}\right)}\nonumber\\
&+\frac{I_{1}\left(q_{0}\right)I_{1}\left(p_{0}+q_{0}\right)}{q_{0}\left(p_{0}+q_{0}\right)}
  -\frac{I_{1}\left(p_{0}\right)I_{1}\left(q_{0}\right)}{p_{0}q_{0}}\Bigg],
\end{align}
where $p_0\in \left\lbrace 2 \pi T n | n\in\mathbb{Z}\right\rbrace$ are understood as the bosonic Matsubara modes and, 
\begin{equation}
I_{n}\left(m\right)\equiv\int_{\boldsymbol{p}}\frac{1}{\left(\boldsymbol{p}^{2}+m^{2}\right)^n}, 
\end{equation}
is a standard integral. This is enough to \emph{almost} factorise the correlator, as a linear combination of two 1-loop sum-integrals. The general 1-loop sum-integral is solved analytically as,
\begin{align} \label{eq:eresult}
  \mathcal{E}^{a}_{m}(t) \equiv &\sumint_P e^{ip_0 t} \frac{p_0^{a}}{P^{2m}}= \frac{T^{4-2m+a}}{\left(2\pi\right)^{2m-a}}
  \frac{\Gamma\left(m-\frac{d}{2}\right)}{\Gamma\left(m\right)}\pi^{\frac{3}{2}}\left(\frac{e^{\gamma_{E}}\bar{\Lambda}^{2}}{4\pi^{2}T^{2}}\right)^{\frac{3-d}{2}}\nonumber\\
    &\times\left[\mathrm{Li}_{2m-a-d}\left(e^{2\pi iTt}\right)+(-1)^{a}\mathrm{Li}_{2m-a-d}\left(e^{-2\pi iTt}\right)\right],
    \end{align}
where $\mathrm{Li}_s(z)$ is the standard polylogarithm. Note that all of the integrals we evaluate are rotationally symmetric. Therefore, we write,
\begin{align}
    \sumint_P& f(P) \equiv T \sum_{p_0} \int_{\mathbf{p}} f(p_0,p)\\
    &\equiv T\lambda(S^{d-1})\left( \frac{e^{\gamma_\mathrm{E}}\bar{\Lambda}^2}{4\pi} \right)^{\frac{3-d}{2}} \int_0^{\infty} \mathrm{d} p \, p^{d-1} \sum_{n\in\mathbb{Z}} f(2\pi Tn,p),\nonumber
\end{align}
where $P=\left(p_0,\mathbf{p}\right)$, $|\mathbf{p}|\equiv p$, $p_0\in \left\lbrace 2 \pi T n | n\in\mathbb{Z}\right\rbrace$ are the bosonic Matsubara modes,
$\displaystyle \int_{\mathbf{p}} \equiv \left(\frac{e^{\gamma_\mathrm{E}}\bar{\Lambda}^2}{4\pi} \right)^{\frac{3-d}{2}} $ 
$\displaystyle \times \int \frac{\mathrm{d}^dp}{(2\pi)^{d}}$,
$\bar{\Lambda}$ is the renormalisation scale, and
$\lambda(S^{d-1})\equiv\left[\Gamma\left(d/2\right)\pi^{d/2}2^{d-1}\right]^{-1}$.
For fermionic Matsubara sum-integral, we use the standard formula, \begin{align}
    \sumint_{\lbrace P\rbrace} e^{ip_0 t} \frac{p_0^{a}}{P^{2m}}=2^{2m-a-d}\mathcal{E}^a_m \left( t/2 \right)-\mathcal{E}^a_m \left( t \right).
\end{align}
In using the IBP identity, we find that there is exactly one non-factorisable contribution which turns out it cannot be fully evaluated analytically. It is only included in one topology, where one contracts a Wilson line with the three-point vertex. For example let's focus on the upper correlator, then the non-factorisable contribution is shown to be proportional to a convolution integral of two 1-loop integrals,
\begin{align}
    \raisebox{-0.44\height}{
\begin{tikzpicture}[scale=0.60*\picSc]
	\doublePlainLoop{2}{0}{2}
	\gluonLine{0}{0}{4}{0}
	\gluonLine{2}{1.9}{2}{0}
	\blobNode{0}{0}{\bigBlobSc}{$E$}
	\blobNode{4}{0}{\bigBlobSc}{$E$}
\end{tikzpicture}} \quad \Bigg|_\mathrm{non-fact.} &
\propto \frac{i}{2}\int_{0}^{t}\mathrm{d}t'\mathcal{E}_{1}^{1}\left(t'\right)\mathcal{E}_{1}^{0}\left(t-t'\right), \label{eq:Ixr}
\end{align}
which we further study by splitting the convolution integral to its non zero mode and its only non vanishing zero mode contribution,
\begin{align}
        \frac{i}{2}&\int_{0}^{t}\mathrm{d}t'\sumint_{PQ}\frac{e^{ip_{0}\left(t-t'\right)}q_{0}e^{iq_{0}t}}{\left(\left(\mathbf{p}+\mathbf{q}\right)^{2}+(p_{0}+q_0)^{2}\right)\left(q^{2}+q_{0}^{2}\right)} \nonumber\\
    &= \frac{i}{2}\int_{0}^{t}\mathrm{d}t'\mathcal{E}_{1}^{1}\left(t'\right)\mathcal{E}_{1}^{0}\left(t-t'\right)|_{\mathrm{p_0 \neq 0}}\\
    &+\frac{t}{2}iT\int_{\mathbf{p}}\sumint_{Q}\frac{q_{0}e^{iq_{0}t}}{\left(\left(\mathbf{p}+\mathbf{q}\right)^{2}+q_0^{2}\right)\left(q^{2}+q_{0}^{2}\right)}\nonumber
    \label{eq:Ixr2},
\end{align}
where we define the zero mode integral as,
\begin{equation}
\mathcal{Z}^{1}_{011} (t) \equiv \sumint_{P} \int_{\mathbf{q}} \frac{p_0 e^{i p_0 t}}{(p_0^2+p^2)(p_0^2+q^2)}.
\end{equation}
Interestingly, under the replacement $t\to 1/T-t$ we find that the non zero mode, shared in all three correlators, is symmetric under the replacement where as, the zero mode is not. The zero mode is proportional to the length of the Wilson line associated with the correlators we study and is their primary distinguishing feature.

We note that the non zero mode integral does not appear to have a closed form solution. We expect it to be divergent at the boundaries i.e at $0$ and $t$. We are able to extract the divergences in closed form as they contribute to the divergence of the correlator which is later reabsorbed in the renormalization of the charge. To achieve this, we have performed a multivariate Taylor expansion. By adding (subtracting) sufficient contributions, we render the integral finite while writing down products of 1-loop integrals that contain the divergence in a closed form under \eqref{eq:eresult}. The remaining integral is now finite and is numerically integrated. More information on the multivariate Taylor expansion alongside with the resulting expressions can be found in Appendix B of \cite{Brambilla:2025xnw}.

We conclude this section by briefly mentioning that the NLO results are not infrared-divergent \cite{Burnier:2008ia,Burnier:2010rp,Binder:2021otw}. This means that the NLO correlator is not sensitive to physics from the soft (Debye mass) scale. Regardless, such a contribution, coming from the Hard Thermal Loop (HTL) resummation of the gluon self energy diagrams has been computed in the literature \cite{Burnier:2010rp}. Alongside our NLO results, this provides a complete $g_s^5$ contribution to the correlator. Therefore, the correlator at order $O(g_s^5)$ reads,
\begin{align}\label{eq:soft}
  \langle& EE\rangle_{I}\Big|_{\text{to order\;} g_s^5}
  = \langle EE\rangle_{I} \\&+  6 N_c T^{-4} \int_{0}^{+\infty} 
  \!\frac{\mathrm{d} \omega}{\pi}\, \rho^{\mathrm{soft}}(\omega) \frac{\cosh\left(\frac{\omega}{2T}-\omega t\right)}{\sinh\left({\frac{\omega}{2T}}\right)} ,\nonumber\label{eq:soft}
\end{align}
where $\langle EE\rangle_{I}$ is the correlator at NLO, whose expression can be found in eq. (3.34) of \cite{Brambilla:2025xnw} and $\rho^{\mathrm{soft}}(\omega)$ is the soft  spectral density and can be read off from \cite{Caron-Huot:2009ncn,Burnier:2010rp}. We find that numerically, the contribution of the soft modes to the correlator is very small.

\section{Results}
Summing all diagrams in eq. \eqref{eq: diagrams} relevant to either the upper or lower correlator, we obtain the nonrenormalised expressions of the upper and lower correlators in $d-$dimensions, which can be found in eq. (3.32) of \cite{Brambilla:2025xnw}. To obtain an expression for the symmetric correlator, we add all diagrams in eq. \eqref{eq: diagrams} as well as the two extra topologies in eq. \eqref{eq: symdiagrams}, paying attention to the different colour factors in each case. Using eq. \eqref{eq:eresult} we expand around $d=3$. We then renormalise the bare coupling, $g_s$, by writing the 1-loop relation of the bare coupling with the renormalised coupling in $\overline{\text{MS}}$-scheme, $ g_{s,\overline{\text{MS}}}$,
\begin{equation}
  g_s^2 = g_{s,\overline{\text{MS}}}^2\left(\bar{\Lambda}\right)\left[1+\frac{g_{s,\overline{\text{MS}}}^2\left(\bar{\Lambda}\right)}{3\varepsilon(4\pi)^2}\left(\frac{e^{\gamma_E}\bar{\Lambda}^2}{4\pi}\right)^{-\varepsilon}\left(2N_f-11N_c\right)\right],
\label{gsMSbar}
\end{equation}
where $\bar{\Lambda}$ is the renormalisation scale. The poles in $\varepsilon$ cancel.

We note that, accounting for the differences in colour factors and at $O(\varepsilon)$ in the $\varepsilon$-expansion, there exist relations between the symmetric correlator and the upper and lower correlator. We find,
\begin{align}
C&\langle EE \rangle_S - \langle EE \rangle_U = -\frac{i}{2\varepsilon} d_A N_c g_s^4  T^{-4}\mathcal{Z}^{1}_{011}(t) + O(\varepsilon) \nonumber\\
&=2\pi T^4\left[\mathrm{Li}^{(1)}_{-3}(e^{2i\pi Tt})-\mathrm{Li}^{(1)}_{-3}(e^{-2i\pi Tt})\right]\varepsilon+O(\varepsilon^2)\nonumber\\
&=  -\frac{3}{\left(2\pi\right)^3} \pi d_A N_c g_s^4\left[\zeta\left(4,Tt\right)-\zeta\left(4,1-Tt\right)\right]+O(\varepsilon),
\label{DiffEESEEU}\\
C&\langle EE \rangle_S - \langle EE \rangle_L = \frac{i}{2\varepsilon} d_A N_c g_s^4  T^{-4}\mathcal{Z}^{1}_{011}(t)  + O(\varepsilon) \nonumber\\
&=  \frac{3}{\left(2\pi\right)^3} \pi d_A N_c g_s^4\left[\zeta\left(4,Tt\right)-\zeta\left(4,1-Tt\right)\right]+O(\varepsilon),
\label{DiffEESEEL}
\end{align}
where $C$ is a colour rescaling factor and $\mathrm{Li}^{(1)}_s(z)$ is the first derivative of the polylogarithm $\mathrm{Li}_s(z)$ with respect to $s$. For example, for the symmetric correlator with $x_{abc} = d_{abc}$, $C = N_c/\left(N_c^2-4\right)$ and for the symmetric correlator with $x_{abc} = f_{abc}$, $C =1/N_c$. The second and third equality are related using $\mathrm{Li}_{s}(e^{2i\pi Tt})+(i)^{2s}\mathrm{Li}_{s}(e^{-2i\pi Tt}) = (2\pi)^si^s\zeta(1-s,Tt)/\Gamma(s)$, where $\zeta\left(s,a\right)$ is the Hurwitz zeta function. The result in eq. \eqref{DiffEESEEU} gives the difference between the fundamental and adjoint correlators first reported in \cite{Scheihing-Hitschfeld:2023tuz}.

It is generally expected that bosonic sum-integrals are symmetric over the thermal circle therefore we expect,  $\left \langle EE\right \rangle_I(t)=\left \langle EE\right \rangle_I\left(1/T-t\right)$. However, this statement is true for correlators constructed from local operators. Here we instead have,  $\left \langle EE\right \rangle_U(t)=\left \langle EE\right \rangle_L\left(1/T-t\right)$, which asserts that the upper and lower correlators never coincide. This allows us to define the asymmetric contribution to the correlator as,
\begin{align}
    \langle EE& \rangle _\mathrm{A} (t) \equiv \frac{1}{2}\left[\left \langle EE \right \rangle_U (t) - \left \langle EE \right \rangle_U (1/T-t)\right]\\  
&= \frac{d_A N_c g_s^4T^{-4}}{2i}  \left(d-1+\frac{2}{d-3}\right) \mathcal{Z}^{1}_{011}(t) \nonumber \\
&=  \frac{3}{\left(2\pi\right)^3} \pi d_A N_c g_s^4\left[\zeta\left(4,Tt\right)-\zeta\left(4,1-Tt\right)\right]+O(\varepsilon). \label{eq:eweasym}\nonumber
\end{align}
At NLO, the asymmetry is associated with the zero modes of the Wilson line.
The zero modes measure the length of the Wilson line, which distinguishes the upper and lower correlators.
Therefore, we expect zero modes to be a major source of the asymmetry also at higher orders. However an all-order proof remains elusive.

We can proceed by plotting the renormalised expression for the correlator. For the renormalised coupling in the $\overline{\text{MS}}$ scheme,  $g_{s,\overline{\text{MS}}}^2 (\bar{\Lambda})$, we use the one-loop running coupling, as this is consistent with our precision. 
We set the renormalisation scale, $\bar{\Lambda}$, to the one-loop optimised scale following \cite{Burnier:2010rp,Brambilla:2022xbd}, 
\begin{equation} \label{eq:pmsscale}
    \bar{\Lambda}_\mathrm{PMS} = 4\pi T \exp \left[-\gamma_E - \frac{N_c-8N_f \ln 2}{2\left(11N_c-2N_f\right)} \right]. 
\end{equation}
We obtain the uncertainty bands by varying this central scale value by a factor of two.

\begin{figure}[ht]
\centering
\includegraphics[width=0.9\columnwidth]{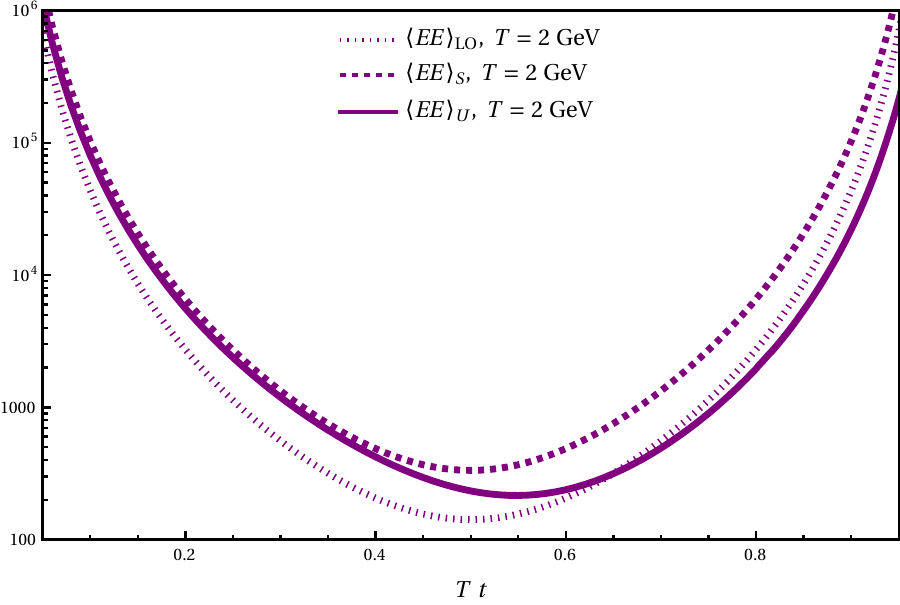}
\caption{The  upper chromoelectric correlator \eqref{defEEU}, the symmetric correlator with $x_{abc}=d_{abc}$ \eqref{defEES} and the LO correlator, $\langle EE \rangle_{\mathrm{LO}}$  as a function of $Tt$ as in \cite{Brambilla:2025xnw}.   
\label{fig:ppmainewe1}}
\end{figure}

In Figure \ref{fig:ppmainewe1}, we plot in logarithmic scale the upper, symmetric and leading-order correlators at $T= 2\,\mathrm{GeV}$ with $N_f= 3$. We have chosen to only portray the central value of the renormalization scale for visual clarity.  We observe that the symmetric correlator is shifted upwards in comparison with LO correlator but both are symmetric over the thermal circle. Interestingly, the minimum of the upper correlator is shifted away from the centre of the thermal circle. This suggests that the correlator is not symmetric over the thermal circle.

\begin{figure}[t]
\centering
\includegraphics[width=0.95\columnwidth]{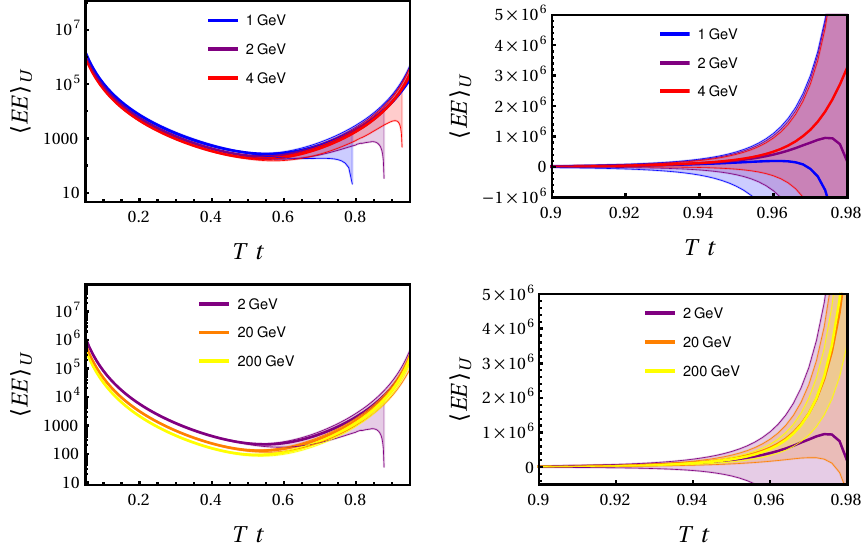}
\caption{The upper chromoelectric correlator \eqref{defEEU} as a function of $Tt$ evaluated at a range of temperatures for $N_c = N_f = 3$, with the renormalisation scale set at $\bar{\Lambda}_{\mathrm{PMS}}$, taken from \cite{Brambilla:2025xnw}.}
\label{fig:ewetemps}
\end{figure}

Figure \ref{fig:ewetemps} illustrates the asymmetry of the upper correlator, with uncertainty bands from scale variation shown for several temperatures.
With our time-independent renormalisation scale, see eq. \eqref{eq:pmsscale}, the region of the thermal circle, where we trust the results grows as the temperature increases. While convergence at large $Tt$ demands very high temperatures, the correlator is already stable at lower $Tt$ for relatively low temperatures.

\begin{figure}[t]
\centering
\includegraphics[width=0.9\columnwidth]{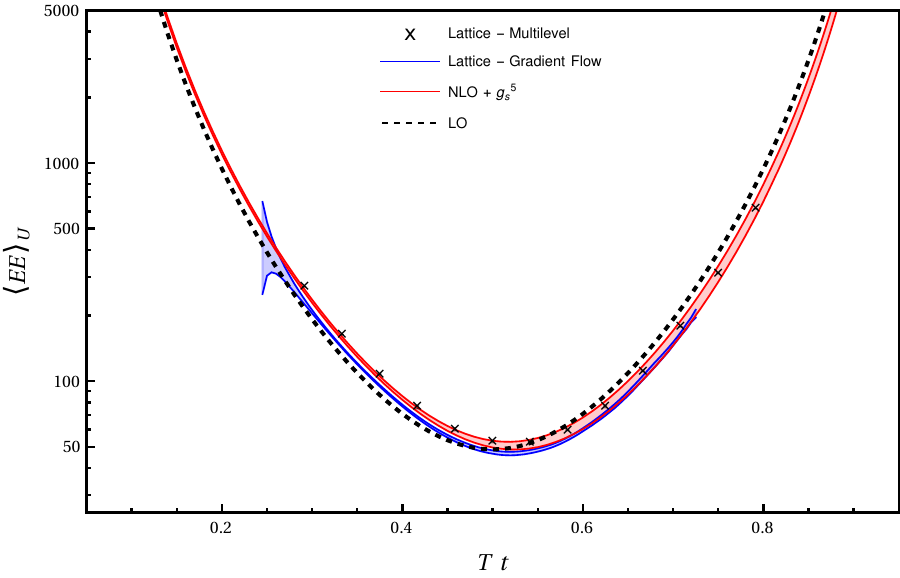}
\caption{Comparison between the lattice data obtained using gradient flow (blue band) and multilevel algorithm (points) of \cite{Brambilla:2025cqy} and the
  $g_s^5$ computation from  this work (red band) for the upper correlator $\langle EE \rangle _U (t)$ at $N_c=3$, $N_f=0$ and $T = 10^4 T_c \approx 3 \times 10^3 \mathrm{GeV}$. Taken from \cite{Brambilla:2025xnw}.
\label{fig:compplot}}
\end{figure}

We have compared our results to a recent lattice study in \cite{Brambilla:2025cqy} using gradient flow \cite{Narayanan:2006rf,Luscher:2009eq,Luscher:2010iy} and the multilevel algorithm \cite{Luscher:2001up}. The results indicate that the upper correlator, responsible for the singlet-to-octet transition in pNRQCD, is asymmetric on the thermal circle. In Fig.~\ref{fig:compplot} we compare our perturbative result, including the $g_s^5$
terms of eq.~\eqref{eq:soft}, with lattice data at
$N_f=0$ and
$T\approx10^4T_c$.
Both show the same functional form, skewed towards the end of the thermal circle. In perturbation theory, uncertainties grow at large 
$Tt$ due to scale choice and the uncertainty of the asymmetric contribution, while in lattice results, the gradient-flow errors increase at small
$Tt$. The errors are both systematic and statistical. Multilevel errors are mainly statistical. The perturbative and lattice results agree well within the corresponding uncertainties. Details on the lattice calculation can be found in \cite{Brambilla:2025cqy}.

\section{Conclusions}
Inspired by quarkonium dynamics, we studied Euclidean correlators of two chromoelectric fields connected by adjoint Wilson lines at finite temperature. Using integration-by-parts, we computed three distinct gauge-invariant correlators to NLO and included soft contributions up to order 
$g_s^5$. One correlator is symmetric under $t\to 1/T-t$ by construction.
While the other two acquire an asymmetry from Wilson line zero modes. Comparison with lattice calculations at very high temperatures shows good agreement.
\section{Acknowledgements}
I thank Nora Brambilla, Saga Säppi and Antonio Vairo for their collaboration and invaluable guidance for the work presented here.




\begin{thebibliography}{00}


\bibitem{Matsui:1986dk}
T.~Matsui and H.~Satz, \emph{{$J/\psi$ Suppression by Quark-Gluon Plasma
  Formation}}, \href{https://doi.org/10.1016/0370-2693(86)91404-8}{\emph{Phys.
  Lett. B} {\bfseries 178} (1986) 416}.

\bibitem{PHENIX:2011img}
{\scshape PHENIX} collaboration, \emph{{$J/\psi$ suppression at forward
  rapidity in Au+Au collisions at $\sqrt{s_{NN}}=200$ GeV}},
  \href{https://doi.org/10.1103/PhysRevC.84.054912}{\emph{Phys. Rev. C}
  {\bfseries 84} (2011) 054912}
  [\href{https://arxiv.org/abs/1103.6269}{{\ttfamily 1103.6269}}].

\bibitem{ALICE:2012jsl}
{\scshape ALICE} collaboration, \emph{{$J/\psi$ suppression at forward rapidity
  in Pb-Pb collisions at $\sqrt{s_{NN}}=2.76$ TeV}},
  \href{https://doi.org/10.1103/PhysRevLett.109.072301}{\emph{Phys. Rev. Lett.}
  {\bfseries 109} (2012) 072301}
  [\href{https://arxiv.org/abs/1202.1383}{{\ttfamily 1202.1383}}].

\bibitem{CMS:2016rpc}
{\scshape CMS} collaboration, \emph{{Suppression of $\Upsilon(1S),
  \Upsilon(2S)$ and $\Upsilon(3S)$ production in PbPb collisions at
  $\sqrt{s_{\rm NN}}$ = 2.76 TeV}},
  \href{https://doi.org/10.1016/j.physletb.2017.04.031}{\emph{Phys. Lett. B}
  {\bfseries 770} (2017) 357}
  [\href{https://arxiv.org/abs/1611.01510}{{\ttfamily 1611.01510}}].

\bibitem{ATLAS:2018hqe}
{\scshape ATLAS} collaboration, \emph{{Prompt and non-prompt $J/\psi $ and
  $\psi (2\mathrm {S})$ suppression at high transverse momentum in
  $5.02~\mathrm {TeV}$ Pb+Pb collisions with the ATLAS experiment}},
  \href{https://doi.org/10.1140/epjc/s10052-018-6219-9}{\emph{Eur. Phys. J. C}
  {\bfseries 78} (2018) 762}
  [\href{https://arxiv.org/abs/1805.04077}{{\ttfamily 1805.04077}}].

\bibitem{STAR:2019fge}
{\scshape STAR} collaboration, \emph{{Measurement of inclusive $J/\psi$
  suppression in Au+Au collisions at $\sqrt{s_{NN}}$ = 200 GeV through the
  dimuon channel at STAR}},
  \href{https://doi.org/10.1016/j.physletb.2019.134917}{\emph{Phys. Lett. B}
  {\bfseries 797} (2019) 134917}
  [\href{https://arxiv.org/abs/1905.13669}{{\ttfamily 1905.13669}}].

\bibitem{Karsch:1987pv}
F.~Karsch, M.T.~Mehr and H.~Satz, \emph{{Color Screening and Deconfinement for
  Bound States of Heavy Quarks}},
  \href{https://doi.org/10.1007/BF01549722}{\emph{Z. Phys. C} {\bfseries 37}
  (1988) 617}.

\bibitem{Kharzeev:1994pz}
D.~Kharzeev and H.~Satz, \emph{{Quarkonium interactions in hadronic matter}},
  \href{https://doi.org/10.1016/0370-2693(94)90604-1}{\emph{Phys. Lett. B}
  {\bfseries 334} (1994) 155}
  [\href{https://arxiv.org/abs/hep-ph/9405414}{{\ttfamily hep-ph/9405414}}].

\bibitem{Xu:1995eb}
X.-M.~Xu, D.~Kharzeev, H.~Satz and X.-N.~Wang, \emph{{$J/\psi$ suppression in
  an equilibrating parton plasma}},
  \href{https://doi.org/10.1103/PhysRevC.53.3051}{\emph{Phys. Rev. C}
  {\bfseries 53} (1996) 3051}
  [\href{https://arxiv.org/abs/hep-ph/9511331}{{\ttfamily hep-ph/9511331}}].

\bibitem{Grandchamp:2001pf}
L.~Grandchamp and R.~Rapp, \emph{{Thermal versus direct $J/\psi$ production in
  ultrarelativistic heavy ion collisions}},
  \href{https://doi.org/10.1016/S0370-2693(01)01311-9}{\emph{Phys. Lett. B}
  {\bfseries 523} (2001) 60}
  [\href{https://arxiv.org/abs/hep-ph/0103124}{{\ttfamily hep-ph/0103124}}].

\bibitem{Grandchamp:2002wp}
L.~Grandchamp and R.~Rapp, \emph{{Charmonium suppression and regeneration from
  SPS to RHIC}},
  \href{https://doi.org/10.1016/S0375-9474(02)01027-8}{\emph{Nucl. Phys. A}
  {\bfseries 709} (2002) 415}
  [\href{https://arxiv.org/abs/hep-ph/0205305}{{\ttfamily hep-ph/0205305}}].

\bibitem{Laine:2006ns}
M.~Laine, O.~Philipsen, P.~Romatschke and M.~Tassler, \emph{{Real-time static
  potential in hot QCD}},
  \href{https://doi.org/10.1088/1126-6708/2007/03/054}{\emph{JHEP} {\bfseries
  03} (2007) 054} [\href{https://arxiv.org/abs/hep-ph/0611300}{{\ttfamily
  hep-ph/0611300}}].

\bibitem{Brambilla:2008cx}
N.~Brambilla, J.~Ghiglieri, A.~Vairo and P.~Petreczky, \emph{{Static
  quark-antiquark pairs at finite temperature}},
  \href{https://doi.org/10.1103/PhysRevD.78.014017}{\emph{Phys. Rev. D}
  {\bfseries 78} (2008) 014017}
  [\href{https://arxiv.org/abs/0804.0993}{{\ttfamily 0804.0993}}].

\bibitem{Brambilla:2011sg}
N.~Brambilla, M.A.~Escobedo, J.~Ghiglieri and A.~Vairo, \emph{{Thermal width
  and gluo-dissociation of quarkonium in pNRQCD}},
  \href{https://doi.org/10.1007/JHEP12(2011)116}{\emph{JHEP} {\bfseries 12}
  (2011) 116} [\href{https://arxiv.org/abs/1109.5826}{{\ttfamily 1109.5826}}].

\bibitem{Brambilla:2013dpa}
N.~Brambilla, M.A.~Escobedo, J.~Ghiglieri and A.~Vairo, \emph{{Thermal width
  and quarkonium dissociation by inelastic parton scattering}},
  \href{https://doi.org/10.1007/JHEP05(2013)130}{\emph{JHEP} {\bfseries 05}
  (2013) 130} [\href{https://arxiv.org/abs/1303.6097}{{\ttfamily 1303.6097}}].

\bibitem{Braun-Munzinger:2000csl}
P.~Braun-Munzinger and J.~Stachel, \emph{{(Non)thermal aspects of charmonium
  production and a new look at $J/\psi$ suppression}},
  \href{https://doi.org/10.1016/S0370-2693(00)00991-6}{\emph{Phys. Lett. B}
  {\bfseries 490} (2000) 196}
  [\href{https://arxiv.org/abs/nucl-th/0007059}{{\ttfamily nucl-th/0007059}}].

\bibitem{Thews:2000rj}
R.L.~Thews, M.~Schroedter and J.~Rafelski, \emph{{Enhanced $J/\psi$ production
  in deconfined quark matter}},
  \href{https://doi.org/10.1103/PhysRevC.63.054905}{\emph{Phys. Rev. C}
  {\bfseries 63} (2001) 054905}
  [\href{https://arxiv.org/abs/hep-ph/0007323}{{\ttfamily hep-ph/0007323}}].

\bibitem{Andronic:2007bi}
A.~Andronic, P.~Braun-Munzinger, K.~Redlich and J.~Stachel, \emph{{Evidence for
  charmonium generation at the phase boundary in ultra-relativistic nuclear
  collisions}},
  \href{https://doi.org/10.1016/j.physletb.2007.07.036}{\emph{Phys. Lett. B}
  {\bfseries 652} (2007) 259}
  [\href{https://arxiv.org/abs/nucl-th/0701079}{{\ttfamily nucl-th/0701079}}].

\bibitem{Zhao:2007hh}
X.~Zhao and R.~Rapp, \emph{{Transverse Momentum Spectra of $J/\psi$ in
  Heavy-Ion Collisions}},
  \href{https://doi.org/10.1016/j.physletb.2008.03.068}{\emph{Phys. Lett. B}
  {\bfseries 664} (2008) 253}
  [\href{https://arxiv.org/abs/0712.2407}{{\ttfamily 0712.2407}}].

\bibitem{Brambilla:2023hkw}
N.~Brambilla, M.A.~Escobedo, A.~Islam, M.~Strickland, A.~Tiwari, A.~Vairo
  et~al., \emph{{Regeneration of bottomonia in an open quantum systems
  approach}}, \href{https://doi.org/10.1103/PhysRevD.108.L011502}{\emph{Phys.
  Rev. D} {\bfseries 108} (2023) L011502}
  [\href{https://arxiv.org/abs/2302.11826}{{\ttfamily 2302.11826}}].


\bibitem{Escobedo:2008sy}
M.A.~Escobedo and J.~Soto, \emph{{Non-relativistic bound states at finite
  temperature (I): The Hydrogen atom}},
  \href{https://doi.org/10.1103/PhysRevA.78.032520}{\emph{Phys. Rev. A}
  {\bfseries 78} (2008) 032520}
  [\href{https://arxiv.org/abs/0804.0691}{{\ttfamily 0804.0691}}].



\bibitem{Brambilla:2010vq}
N.~Brambilla, M.A.~Escobedo, J.~Ghiglieri, J.~Soto and A.~Vairo, \emph{{Heavy
  Quarkonium in a weakly-coupled quark-gluon plasma below the melting
  temperature}}, \href{https://doi.org/10.1007/JHEP09(2010)038}{\emph{JHEP}
  {\bfseries 09} (2010) 038} [\href{https://arxiv.org/abs/1007.4156}{{\ttfamily
  1007.4156}}].

\bibitem{Brambilla:2016wgg}
N.~Brambilla, M.A.~Escobedo, J.~Soto and A.~Vairo, \emph{{Quarkonium
  suppression in heavy-ion collisions: an open quantum system approach}},
  \href{https://doi.org/10.1103/PhysRevD.96.034021}{\emph{Phys. Rev. D}
  {\bfseries 96} (2017) 034021}
  [\href{https://arxiv.org/abs/1612.07248}{{\ttfamily 1612.07248}}].

\bibitem{Brambilla:2025xnw}N.~Brambilla, P.~Panayiotou, S.~Säppi and A.~Vairo, 
\emph{{The chromoelectric adjoint correlators in Euclidean space at next-to-leading order}},
\href{https://doi.org/10.1007/JHEP08(2025)219}{\emph{ JHEP} {\bfseries 08} (2025) 219} [\href{https://arxiv.org/abs/2505.16604}{{\ttfamily
  2505.16604}}] 


\bibitem{Binder:2021otw}
T.~Binder, K.~Mukaida, B.~Scheihing-Hitschfeld and X.~Yao, \emph{{Non-Abelian
  electric field correlator at NLO for dark matter relic abundance and
  quarkonium transport}},
  \href{https://doi.org/10.1007/JHEP01(2022)137}{\emph{JHEP} {\bfseries 01}
  (2022) 137} [\href{https://arxiv.org/abs/2107.03945}{{\ttfamily
  2107.03945}}].

\bibitem{Biondini:2023zcz}
S.~Biondini, N.~Brambilla, G.~Qerimi and A.~Vairo, \emph{{Effective field
  theories for dark matter pairs in the early universe: cross sections and
  widths}}, \href{https://doi.org/10.1007/JHEP07(2023)006}{\emph{JHEP}
  {\bfseries 07} (2023) 006}
  [\href{https://arxiv.org/abs/2304.00113}{{\ttfamily 2304.00113}}].

\bibitem{Eller:2019spw}
A.M.~Eller, J.~Ghiglieri and G.D.~Moore, \emph{{Thermal Heavy Quark Self-Energy
  from Euclidean Correlators}},
  \href{https://doi.org/10.1103/PhysRevD.99.094042}{\emph{Phys. Rev. D}
  {\bfseries 99} (2019) 094042}
  [\href{https://arxiv.org/abs/1903.08064}{{\ttfamily 1903.08064}}].

\bibitem{Scheihing-Hitschfeld:2023tuz}
B.~Scheihing-Hitschfeld and X.~Yao, \emph{{Real time quarkonium transport
  coefficients in open quantum systems from Euclidean QCD}},
  \href{https://doi.org/10.1103/PhysRevD.108.054024}{\emph{Phys. Rev. D}
  {\bfseries 108} (2023) 054024}
  [\href{https://arxiv.org/abs/2306.13127}{{\ttfamily 2306.13127}}].



\bibitem{Caron-Huot:2009ncn}
S.~Caron-Huot, M.~Laine and G.D.~Moore, \emph{{A Way to estimate the heavy
  quark thermalization rate from the lattice}},
  \href{https://doi.org/10.1088/1126-6708/2009/04/053}{\emph{JHEP} {\bfseries
  04} (2009) 053} [\href{https://arxiv.org/abs/0901.1195}{{\ttfamily
  0901.1195}}].

\bibitem{Burnier:2010rp}
Y.~Burnier, M.~Laine, J.~Langelage and L.~Mether, \emph{{Colour-electric
  spectral function at next-to-leading order}},
  \href{https://doi.org/10.1007/JHEP08(2010)094}{\emph{JHEP} {\bfseries 08}
  (2010) 094} [\href{https://arxiv.org/abs/1006.0867}{{\ttfamily 1006.0867}}].

\bibitem{Brambilla:2025cqy}
{\scshape TUMQCD} collaboration, \emph{{Lattice study of correlators of
  chromoelectric fields for heavy quarkonium dynamics in the quark-gluon
  plasma}},  \href{https://arxiv.org/abs/2505.16603}{{\ttfamily 2505.16603}}.


\bibitem{Brambilla:2017zei}
N.~Brambilla, M.A.~Escobedo, J.~Soto and A.~Vairo, \emph{{Heavy quarkonium
  suppression in a fireball}},
  \href{https://doi.org/10.1103/PhysRevD.97.074009}{\emph{Phys. Rev. D}
  {\bfseries 97} (2018) 074009}
  [\href{https://arxiv.org/abs/1711.04515}{{\ttfamily 1711.04515}}].
  
\bibitem{Brambilla:QuarkoniumTransportCoefficients}
N.~Brambilla, M.A.~Escobedo, A.~Islam, M.~Strickland, A.~Vairo and
  P.~Vander~Griend, \emph{Anatomy of quarkonium transport coefficients, {\rm in
  preparation}},  \href{https://arxiv.org/abs/TUM-EFT 191/24,
  FERMILAB-PUB-24-0451-T.}{{\ttfamily TUM-EFT 191/24, FERMILAB-PUB-24-0451-T.}}

\bibitem{Bellac:2011kqa}
M.~Le~Bellac, "Thermal Field Theory", \emph{Cambridge}, UK: \emph{Univ. Pr. (2011), 256 p.}


\bibitem{Davydychev:2022dcw}
A.I.~Davydychev and Y.~Schr\"oder, \emph{{Recursion-free solution for two-loop
  vacuum integrals with \textquotedblleft{}collinear\textquotedblright{}
  masses}}, \href{https://doi.org/10.1007/JHEP12(2022)047}{\emph{JHEP}
  {\bfseries 12} (2022) 047}
  [\href{https://arxiv.org/abs/2210.10593}{{\ttfamily 2210.10593}}].

\bibitem{Davydychev:2023jto}
A.I.~Davydychev, P.~Navarrete and Y.~Schr\"oder, \emph{{Factorizing two-loop
  vacuum sum-integrals}},
  \href{https://doi.org/10.1007/JHEP02(2024)104}{\emph{JHEP} {\bfseries 02}
  (2024) 104} [\href{https://arxiv.org/abs/2312.17367}{{\ttfamily
  2312.17367}}].

\bibitem{Burnier:2008ia}
Y.~Burnier, M.~Laine and M.~Vepsalainen, \emph{{Heavy quark medium polarization
  at next-to-leading order}},
  \href{https://doi.org/10.1088/1126-6708/2009/02/008}{\emph{JHEP} {\bfseries
  02} (2009) 008} [\href{https://arxiv.org/abs/0812.2105}{{\ttfamily
  0812.2105}}].
  
\bibitem{Brambilla:2022xbd}
{\scshape TUMQCD} collaboration, \emph{{Heavy quark diffusion coefficient with
  gradient flow}},
  \href{https://doi.org/10.1103/PhysRevD.107.054508}{\emph{Phys. Rev. D}
  {\bfseries 107} (2023) 054508}
  [\href{https://arxiv.org/abs/2206.02861}{{\ttfamily 2206.02861}}].


\bibitem{Narayanan:2006rf}
R.~Narayanan and H.~Neuberger, \emph{{Infinite N phase transitions in continuum
  Wilson loop operators}},
  \href{https://doi.org/10.1088/1126-6708/2006/03/064}{\emph{JHEP} {\bfseries
  03} (2006) 064} [\href{https://arxiv.org/abs/hep-th/0601210}{{\ttfamily
  hep-th/0601210}}].

\bibitem{Luscher:2009eq}
M.~L{\"u}scher, \emph{{Trivializing maps, the Wilson flow and the HMC
  algorithm}}, \href{https://doi.org/10.1007/s00220-009-0953-7}{\emph{Commun.
  Math. Phys.} {\bfseries 293} (2010) 899}
  [\href{https://arxiv.org/abs/0907.5491}{{\ttfamily 0907.5491}}].

\bibitem{Luscher:2010iy}
M.~L\"uscher, \emph{{Properties and uses of the Wilson flow in lattice QCD}},
  \href{https://doi.org/10.1007/JHEP08(2010)071}{\emph{JHEP} {\bfseries 08}
  (2010) 071} [\href{https://arxiv.org/abs/1006.4518}{{\ttfamily 1006.4518}}].

\bibitem{Luscher:2001up}
M.~L{\"u}scher and P.~Weisz, \emph{{Locality and exponential error reduction in
  numerical lattice gauge theory}},
  \href{https://doi.org/10.1088/1126-6708/2001/09/010}{\emph{JHEP} {\bfseries
  09} (2001) 010} [\href{https://arxiv.org/abs/hep-lat/0108014}{{\ttfamily
  hep-lat/0108014}}].



\end{thebibliography}



\end{document}